\documentclass[twocolumn,english,aps,prl,floatfix,superscriptaddress]{revtex4-1}
\usepackage[utf8]{luainputenc}
\usepackage{float}
\usepackage{amsmath}
\usepackage{amssymb}
\usepackage{stackrel}
\usepackage{graphicx}
\usepackage{esint}

\makeatletter
\usepackage{babel}

\makeatother

\usepackage{babel}
\begin{document}
\title{Quantum oscillations and criticality in a fermionic and bosonic dimer
model for the cuprates}
\author{Garry Goldstein}
\affiliation{TCM Group, Cavendish Laboratory, University of Cambridge, J. J. Thomson
Avenue, Cambridge CB3 0HE, United Kingdom}
\author{Claudio Chamon}
\affiliation{Department of Physics, Boston University, Boston, Massachusetts 02215,
USA}
\author{Claudio Castelnovo}
\affiliation{TCM Group, Cavendish Laboratory, University of Cambridge, J. J. Thomson
Avenue, Cambridge CB3 0HE, United Kingdom}
\begin{abstract}
We study quantum oscillations for a system of fermionic and bosonic
dimers and compare the results to those experimentally observed in
the cuprate superconductors in their underdoped regime. We argue that
the charge carriers obey the Onsager quantization condition and quantum
oscillations take on a Lifshitz-Kosevich form. We obtain the effective
mass and find good qualitative agreement with experiments if we tune
the model to the point where the observed mass divergence at optimum
doping is associated to a van Hove singularity at which four free-dimer
Fermi pockets touch pairwise in the interior of the Brillouin zone.
The same van Hove singularity leads to a maximum in the d-wave superconducting
pairing amplitude when anti-ferromagnetic interactions are included.
Our combined results therefore suggest that a quantum critical point
separating the underdoped and overdoped regimes is marked by the location
of the van Hove saddle point in the fermionic dimer dispersion. 
\end{abstract}
\maketitle

\section{\label{sec:Introduction}Introduction }

Recent experiments suggest that the pseudogap phase of the high temperature
cuprate superconductors can be described in terms of a vanilla Fermi
liquid with an anomalously low quasiparticle density~\cite{Mirzaei2013,Chan2014,Lebouf2007,Sebastian2011}.
In particular, the observation of quantum oscillations in underdoped
cuprates~\cite{Sebastian2011,Leyraud2007,Bangura2008,Jaudet2008,Sebastian2010,Singleton2010,Ramshaw2015,Barisic2013}
with frequency between $500$ and $600$~T indicates the existence
of a Fermi surface with area $\sim p/8$ (where $p$ is the doping).
What is most convincing evidence of nearly free quasiparticles obeying
Fermi-Dirac statistics is the striking resemblance between the amplitude
of the oscillations as a function of temperature and that predicted
by the Lifshitz-Kosevich formula~\cite{Sebastian2011} ${A(T)}/{A(0)}={\pi\eta}/{\sinh(\pi\eta)}$,
where $\eta=2\pi k_{B}Tm^{*}/\hbar eB$ and the effective mass $m^{*}$
is the only parameter used to fit experiments over a wide range of
temperatures. It is therefore imperative that any candidate model
for the cuprates be capable of explaining these features.

\begin{figure}
\begin{centering}
\includegraphics[width=1\columnwidth]{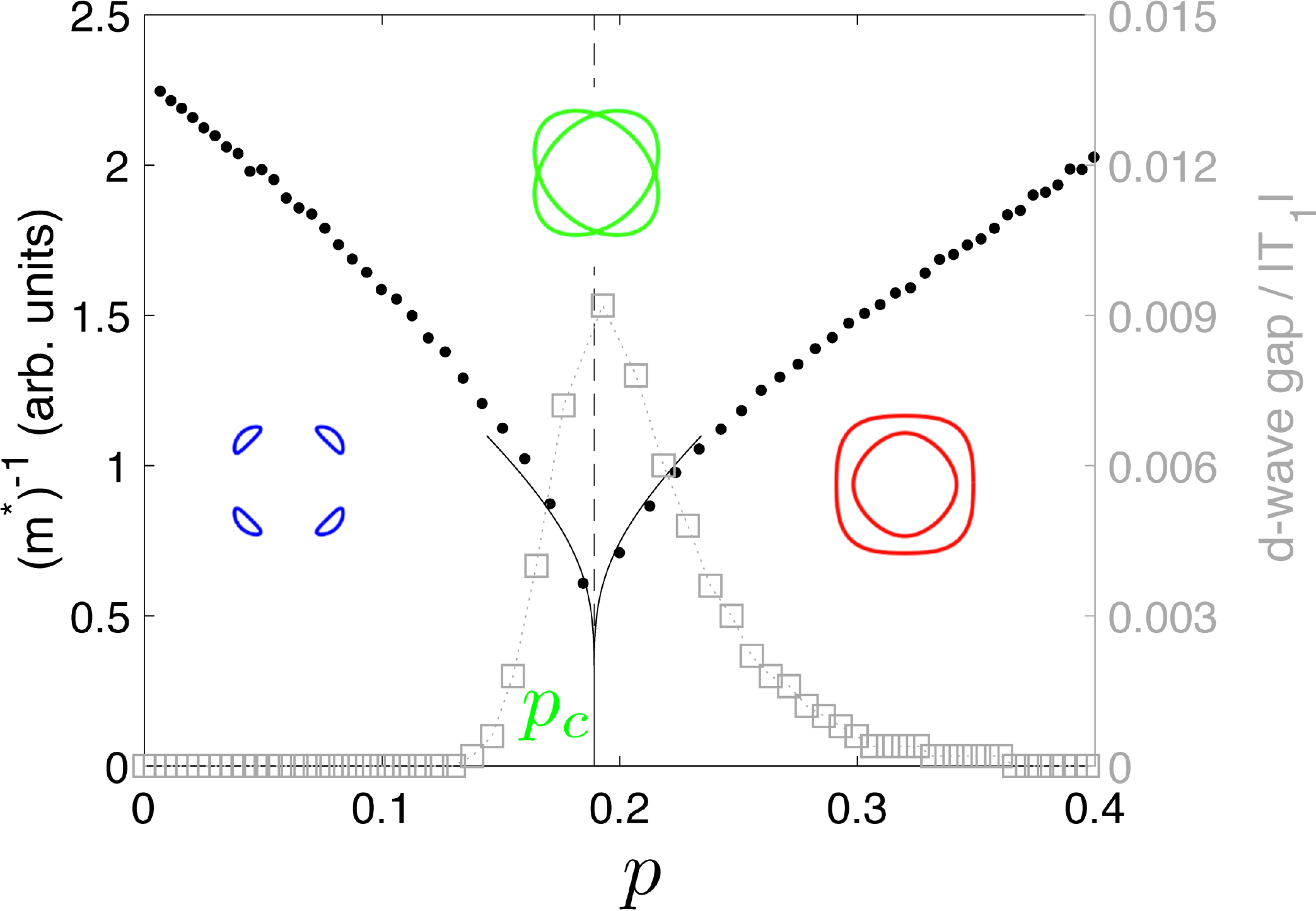} 
\par\end{centering}
\caption{\label{fig-data} (color online) Inverse quasiparticle mass (circles,
left axis) and d-wave superconducting gap (squares, right axis) as
a function of the doping $p$ near the van Hove singularity at $p_{c}$
(vertical dashed line), for $T_{2}/T_{1}=-0.8$, $T_{3}/T_{1}=0.5$
and $J/T_{1}=1.0$. The thin line is the perturbative analytical solution
for the inverse mass near the van Hove singularity, as discussed in
the text. The data synthesizes our theoretical proposal of a quantum
critical point near optimum doping, separating two regimes where the
Fermi surface for the fermionic dimers changes between the two topologies
shown in blue and red. The critical point is marked by the van Hove
singularity, with Fermi surface topology depicted in green.}
\end{figure}
A model of fermionic and bosonic (FB) quantum dimers has recently
been proposed as a candidate for describing the physics of the underdoped
cuprates~\cite{Punk2015,Chowdhury2016,Patel2016,Goldstein2017,Feldelmeier2017,Huber2017}.
The FB dimer model contains spinless bosonic dimers, representing
a valence bond between two neighboring spins, and spin-1/2 fermionic
dimers, representing a hole delocalized between two sites. By condensing
the bosonic dimers, one obtains a tractable mean field effective Hamiltonian
for the fermionic dimers that captures well the emergence of \textit{d}-wave
superconductivity when the Fermi surface of the dimers exhibits appropriate
pockets~\cite{Goldstein2017}.

In this paper we study quantum oscillations in the FB quantum dimer
model, and compare our results with the behavior experimentally observed
in the cuprate superconductors in the underdoped regime. We remark
that, although we concentrate on the FB dimer model, the results here
presented should apply more generally to systems with degrees of freedom
sitting on the bonds, for instance multiorbital models of the cuprates
that include the oxygen sites~\cite{Emery1987}.

In a regime where the magnetic length and the size of the quasiclassical
wavepacket is much larger than the lattice spacing, we argue that
the fermionic dimers behave as free quasiparticles and undergo semiclassical
oscillations under the effect of a magnetic field. These oscillations
obey the Onsager quantization condition and the standard Lifshitz-Kosevich
form, dictated by the minimal coupling of the gauge field to the quasiparticles.

We compute the effective mass of the quasiparticles and find that
it is in good agreement with experiments if we posit that the observed
divergence of the mass at optimum doping is associated to a van Hove
singularity where the dimer pockets merge in the bulk of the Brillouin
zone. Consistently, we find that the superconducting gap and critical
temperature are maximal at the value of doping where the Fermi surface
topology changes, due to the enhanced density of states at the singularity.

Figure~\ref{fig-data} presents results for the effective mass and
superconducting order parameter as functions of doping. This data
summarizes our proposal of a quantum critical point near optimum doping,
separating two regimes where the Fermi surfaces for the fermionic
dimers have different topology, as depicted in the figure. The quantum
critical point corresponds to a van Hove singularity \textit{inside}
the Brillouin zone, \textit{not} at its boundary.

Together with earlier work~\cite{Punk2015,Chowdhury2016,Patel2016,Goldstein2017,Feldelmeier2017,Huber2017},
our results make a substantial contribution to highlight the suggestive
similarity between the behavior of the FB dimer model and the physics
of underdoped cuprates near the superconducting dome. This is most
remarkable given the relative simplicity of the effective dimer description.
Whereas the FB dimer model comes with a number of free parameters
that are difficult to fix from first principles, our work imposes
strong limitations on the range of these parameters where the behavior
of the model compares well with experiments. This brings us within
reach of critically testing the validity and limits of this model
to describe the behavior of underdoped cuprates.

%%%%%%%%%%%%%%%%%%%%%%%%%%%%%%%%%%%%%%%%%%%%%%%%%%%%%%%%%%%%%%%%%%%%%%%

\section{\label{sec:The-effective-model}The effective model}

\begin{figure}
\begin{centering}
\includegraphics[scale=0.25]{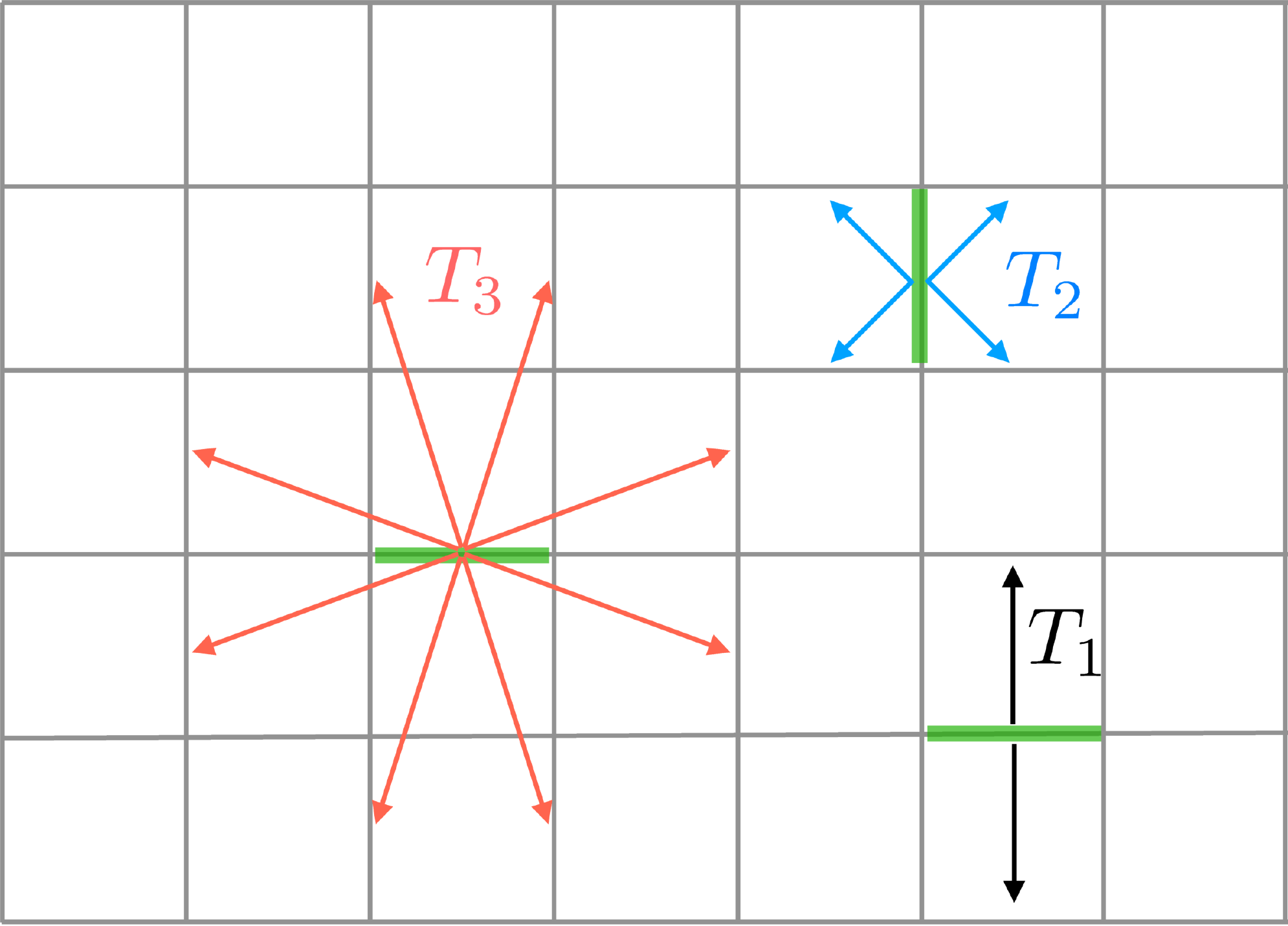} 
\par\end{centering}
\caption{\label{fig:hoppings} (color online) %The effective hoppings of the fermionic dimers upon condensation of 
%the bosonic ones.
The FB quantum dimer model of Ref.~\protect\protect\protect\onlinecite{Punk2015}
contains dimers on the bonds of the square lattice. Condensing the
bosonic dimers leads to a theory of free fermionic dimers, Eq.~\eqref{eq:Main_Hamiltonian},
with effective hoppings $T_{1,2,3}$ that encode both the bare coupling
constants and the expectation values of bilinears in the bosonic dimers,
as presented in Ref.~\protect\protect\protect\onlinecite{Goldstein2017}.
The lattice of bonds contains two sublattices, corresponding to the
vertical and horizontal bonds. The hopping amplitudes $T_{2,3}$ moves
dimers between the two sublattices, while hopping amplitude $T_{1}$
breaks chiral symmetry.}
\end{figure}

In our study of quantum oscillations, we consider the mean field description
presented in Ref.~\onlinecite{Goldstein2017} of the FB dimer model
introduced in Ref.~\onlinecite{Punk2015} to describe the pseudogap
phase of the underdoped cuprates. Substantial progress in understanding
the fermionic component of the theory can be made using the mean field
Hamiltonian obtained by condensing bosonic dimer bilinears, which
renormalize the effective hopping amplitudes for the remaining fermionic
dimers (illustrated pictorially in Fig.~\ref{fig:hoppings}). The
approach is phenomenological, in that we do not compute these amplitudes
microscopically, but instead we treat them as free fitting parameters
$T_{1,2,3}$.

%%%%%%%%%%%%%%%%%%%%%%%%%%%%%%%%%%%%%%
%%%%%%%%%%%%%%%%%%%%%%%%%%%%%%%%%%%%%%

The fermionic mean field Hamiltonian reads~\cite{Goldstein2017}:
\begin{align}
H_{F\bar{B}}= & -T_{1}\;\sum_{i}\sum_{\sigma}\left(c_{i+\hat{y},\hat{x},\sigma}^{\dagger}c_{i,\hat{x},\sigma}^{\;}+c_{i+\hat{x},\hat{y},\sigma}^{\dagger}c_{i,\hat{y},\sigma}^{\;}\right)+{\rm {H.c.}\nonumber}\\
 & -T_{2}\;\sum_{i}\sum_{\sigma}\sum_{v\in V_{2}}c_{i+v,\hat{y},\sigma}^{\dagger}c_{i,\hat{x},\sigma}^{\;}+{\rm {H.c.}\nonumber}\\
 & -T_{3}\;\sum_{i}\sum_{\sigma}\sum_{v\in V_{3}}c_{i+v,\hat{y},\sigma}^{\dagger}c_{i,\hat{x},\sigma}^{\;}+{\rm {H.c.}\nonumber}\\
 & -\mu\;\sum_{i}\sum_{\sigma}\left(c_{i,\hat{x},\sigma}^{\dagger}c_{i,\hat{x},\sigma}^{\;}+c_{i,\hat{y},\sigma}^{\dagger}c_{i,\hat{y},\sigma}^{\;}\right)%\nonumber%
%+{\rm {H.c.}}
\,.\label{eq:Main_Hamiltonian}
\end{align}
The operator $c_{i,\eta,\sigma}$ annihilates a fermion with spin
$\sigma$ on the bond $(i,i+\eta)$, which is horizontal for $\eta=\hat{x}$
or vertical for $\eta=\hat{y}$. Notice that $T_{1}$ hops the fermionic
dimers between parallel bonds, while $T_{2,3}$ flip the dimers from
horizontal to vertical and \textit{vice versa}. We define (in momentum
space) the spinor that encodes the horizontal and vertical flavors
as $\psi_{\vec{k},\sigma}^{\dagger}=(c_{\vec{k},\hat{y},\sigma}^{\dagger},c_{\vec{k},\hat{x},\sigma}^{\dagger})$
and~\cite{Goldstein2017}: 
\begin{equation}
H_{F\bar{B}}=\sum_{\vec{k},\sigma}\psi_{\vec{k},\sigma}^{\dagger}\,\begin{pmatrix}\xi_{\vec{k}}^{x} & \gamma_{\vec{k}}\\
\gamma_{\vec{k}}^{*} & \xi_{\vec{k}}^{y}
\end{pmatrix}\;\psi_{\vec{k},\sigma}\;,\label{eq:H_k}
\end{equation}
where: 
\begin{align*}
\xi_{\vec{k}}^{x,y} & =-\mu-2\,T_{1}\;\cos k_{x,y}\;\\
\gamma_{\vec{k}} & =4\,e^{i(k_{y}-k_{x})/2}\;\left(T_{2}\;\cos\frac{k_{x}}{2}\cos\frac{k_{y}}{2}\right.\\
 & \left.+T_{3}\;\cos\frac{3k_{x}}{2}\cos\frac{k_{y}}{2}+T_{3}\;\cos\frac{k_{x}}{2}\cos\frac{3k_{y}}{2}\right).
\end{align*}
The eigenvalues are given by $E_{\pm,\vec{k}}=\xi_{\vec{k}}\pm\sqrt{\eta_{\vec{k}}^{2}+|\gamma_{\vec{k}}|^{2}}$,
where $\xi_{\vec{k}}=(\xi_{\vec{k}}^{x}+\xi_{\vec{k}}^{y})/2$ and
$\eta_{\vec{k}}=(\xi_{\vec{k}}^{x}-\xi_{\vec{k}}^{y})/2$. The lower
band $E_{-,\vec{k}}$ will be partially occupied upon hole doping,
with concentration $p$. %The Hamiltonian Eq.~\eqref{eq:H_k} has
%four-fold rotational symmetry, $(k_{x},k_{y})\rightarrow(k_{y},-k_{x})$,
%and reflection symmetry about the two axis, 
%$(k_{x},k_{y})\rightarrow(\pm k_{y},\mp k_{x})$.

We shall rescale the Hamiltonian and study $H_{F\bar{B}}/|T_{1}|$,
i.e., work in energy units of $|T_{1}|=1$. We proceed with our investigation
of the model by analyzing its properties as a function of the dimensionless
ratios $T_{2}/|T_{1}|$ and $T_{3}/|T_{1}|$, as well as the doping
$p$ (controlled by the chemical potential $\mu$). The essence of
our approach is to determine the space of parameters of the system
where it matches the phenomenology of the cuprates. For instance,
in Ref.~\onlinecite{Goldstein2017} it was found that the region
in the two dimensional parameter space exhibiting four small Fermi
pockets largely overlapped with the region where \textit{d}-wave superconductivity
existed, when the anti-ferromagnetic coupling $J$ of the $t-J$ model
was included. We note that the state with the lower band fully occupied
corresponds to an unphysical doping $p=2$; however, the physics discussed
in this paper pertains to sensibly small values of the hole doping
$p$ where one can expect the dimer representation to be valid.

\section{\textup{\label{sec:Quantum-Oscillations}Quantum Oscillations}}

Oscillations of magnetoresistence reflect how a system responds to
an applied magnetic field, which always couples minimally to the physical
constituents of the system, i.e., electrons. The fermionic dimers
are not the elementary constituents; they are emerging particles,
and therefore the case for quantum oscillations requires more care.

The Hamiltonian Eq.~\eqref{eq:Main_Hamiltonian} is obtained from
a mean field approximation of an interacting FB dimer model, which
in turn is an effective projection of a microscopic system, such as
the Hubbard model, onto a subspace of dimers. Thus we are faced with
the problem of logically justifying that the mean field Hamiltonian
does capture quantum oscillations of the underlying physical system.

The justification for minimally coupling the dimers to the external
magnetic field hinges on the fact that we restrict our analysis to
the case when the dimer size (set by the lattice spacing) is much
smaller than both the magnetic length and the size of the wavepacket.
In other words, in this regime one cannot resolve the non-elementary
nature of the fermionic dimers. Hence, quantum oscillations in the
FB dimer model are described by those of charged quasiparticles with
dynamics governed by Eq.~\eqref{eq:Main_Hamiltonian} upon shifting
$\vec{k}$ by the gauge potential. In this regime, it is therefore
reasonable to expect the quantum oscillations to satisfy the Onsager
quantization condition as well as the Lifshitz-Kosevich formula.

We note that the conventional expectation for the charge of fermionic
dimers in the FB model is $+e$~\cite{Punk2015}. This sign is consistent
with Hall coefficient measurements at high temperature. However, the
data show a change of sign of the carriers at low temperatures~\cite{Lebouf2007,Leboeuf2011}.
Understanding this phenomenon is beyond the scope of the present paper,
but an explanation may be possible within the FB dimer model if one
accounts for phase factors in the wave function of the bosonic dimers
in the presence of sufficiently large magnetic fields.

One of the salient features of the mean field model governed by Eq.~\eqref{eq:Main_Hamiltonian}
is a region in parameter space $T_{1,2,3}$ where the dispersion exhibits
pockets near the $(\pm\pi/2,\pm\pi/2)$ points. The period of oscillations
depends on the size of the Fermi surfaces, with each disconnected
surface contributing its own frequency. The presence of four identical
Fermi pockets of size $p/8$ (the factor of 2 due to spin) is consistent
with the experimental data~\cite{Sebastian2011,Singleton2010,Ramshaw2015}
in the doping range of 10\%-16\%, with the $p/8$ result being a nearly
ideal intercept at a doping of $\sim$ 13\%. The slope of the $p/8$
curve is higher than the slope of the experimental data with the frequency
of oscillations between 500 and 600 Tesla in the doping range of 10\%-16\%;
however the discrepancy is small (see Appendix \ref{sec:Frequency-of-the}).

Experimentally, the oscillations become less well defined as one approaches
optimal doping. In addition, experimental measurements show that the
effective quasiparticle mass increases as the density increases towards
the optimal doping value, with the extrapolation suggesting a divergence.
This is consistent with the presence of a Fermi surface singularity,
where quantum oscillations are suppressed because of the corresponding
enhancement in the density of states and the residual interactions,
not captured by mean field theory, lead to scattering and departure
from the Lifshitz-Kosevich formula.

\begin{figure}
\begin{centering}
\includegraphics[scale=0.4]{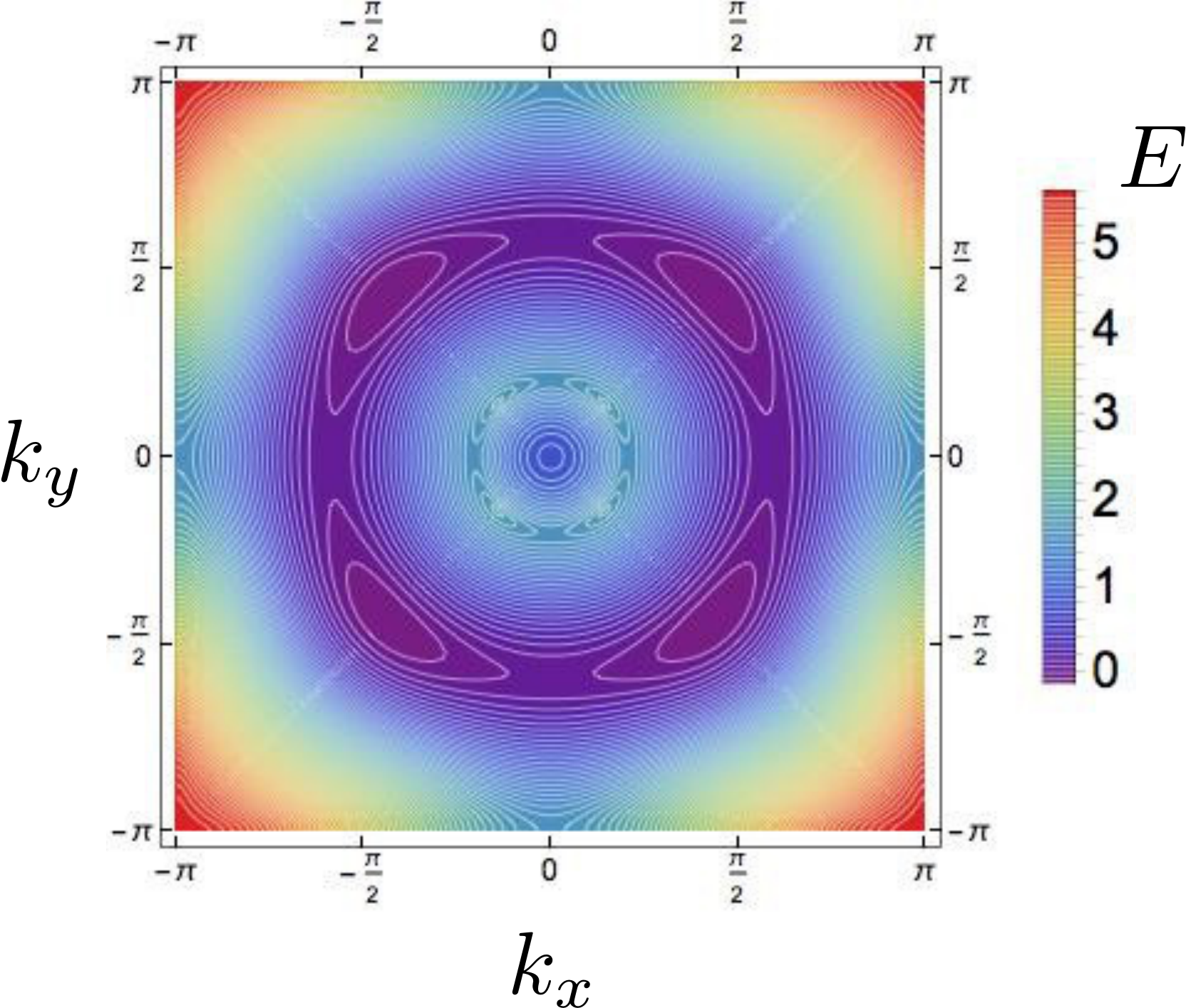} \\
 \vspace{0.9cm}
 \hspace{-1.2cm}\includegraphics[scale=0.4]{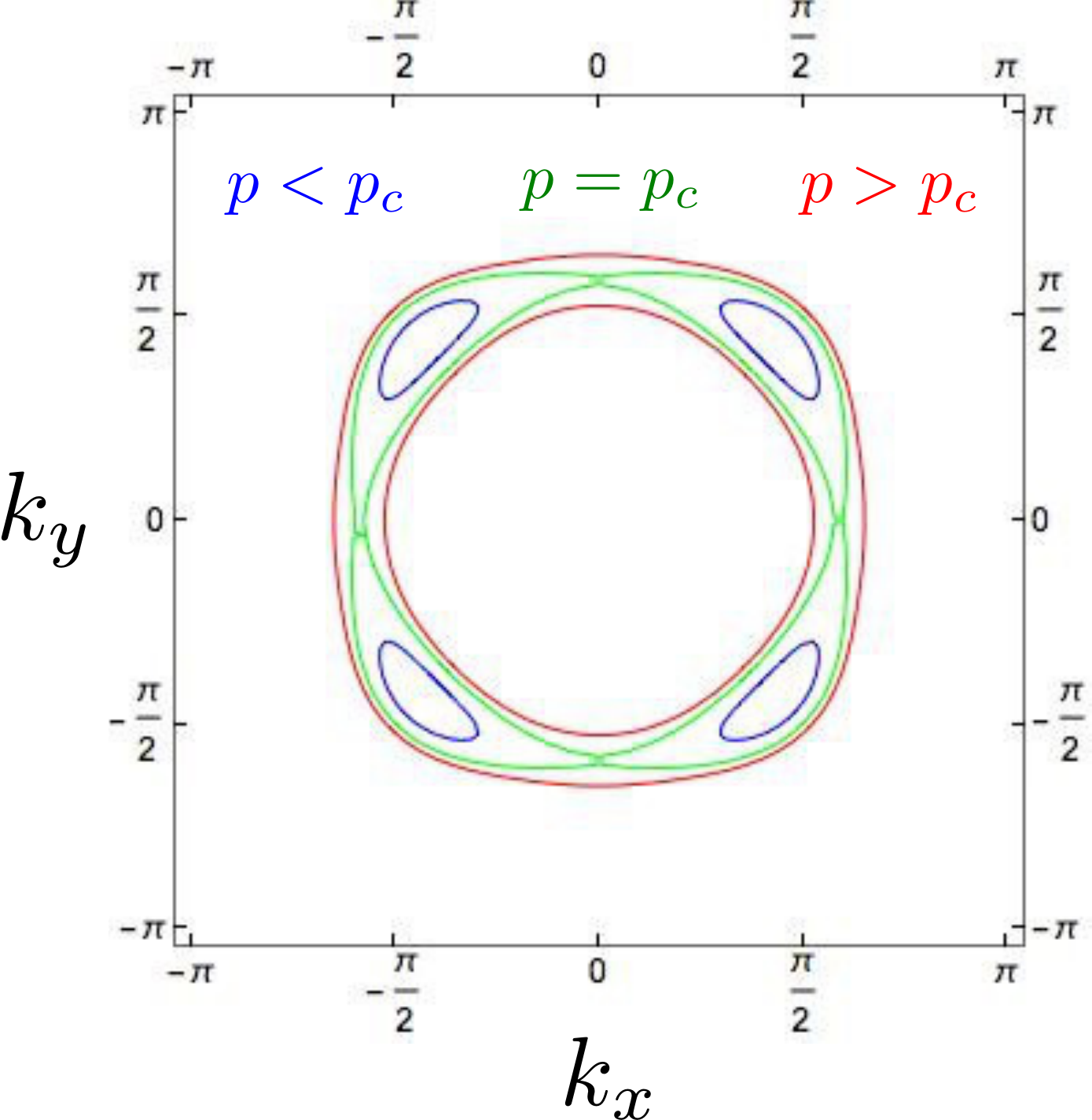} 
\par\end{centering}
\caption{\label{fig-vanHove} (color online) Dispersion for the effective model
of fermionic dimers for $T_{2}/T_{1}=-0.8$ and $T_{3}/T_{1}=0.5$.
Top panel: constant energy surfaces, with $E=0$ measured from the
bottom of the band near $(\pm\pi/2,\pm\pi/2)$. Bottom panel: Fermi
surfaces corresponding to doping levels slightly below, at, and slightly
above optimum doping $p_{c}$, where saddle points in the energy dispersion
occur near $(0,\pm\pi/2)$ and $(\pm\pi/2,0)$.}
\end{figure}

Here we explore the possibility that this suppression of quantum oscillations
near optimal doping corresponds to a new type of van Hove singularity
for the cuprates where the four pockets merge in the bulk of the Brillouin
zone, morphing into two Fermi surfaces with one sheathing the other
(as illustrated in the bottom panel of Fig.~\ref{fig-vanHove}).
Notice that this singularity is different in nature with respect to
the ones previously studied in the context of underdoped cuprates~\cite{Emery1987}
in two main aspects: it does not arise from the competition with an
ordering instability (e.g., CDW), and it takes place away from the
Brillouin zone boundary.

For our proposed scenario to occur, we ought to find a region in parameter
space of the mean field model where: (i) the dispersion exhibits four
pockets; (ii) the targeted van Hove singularity occurs near $p=0.2$
(say within $\pm0.02$), and (iii) the leading superconducting instability
in the presence of interactions is \textit{d}-wave.

\begin{figure}
\begin{centering}
\includegraphics[width=0.95\columnwidth]{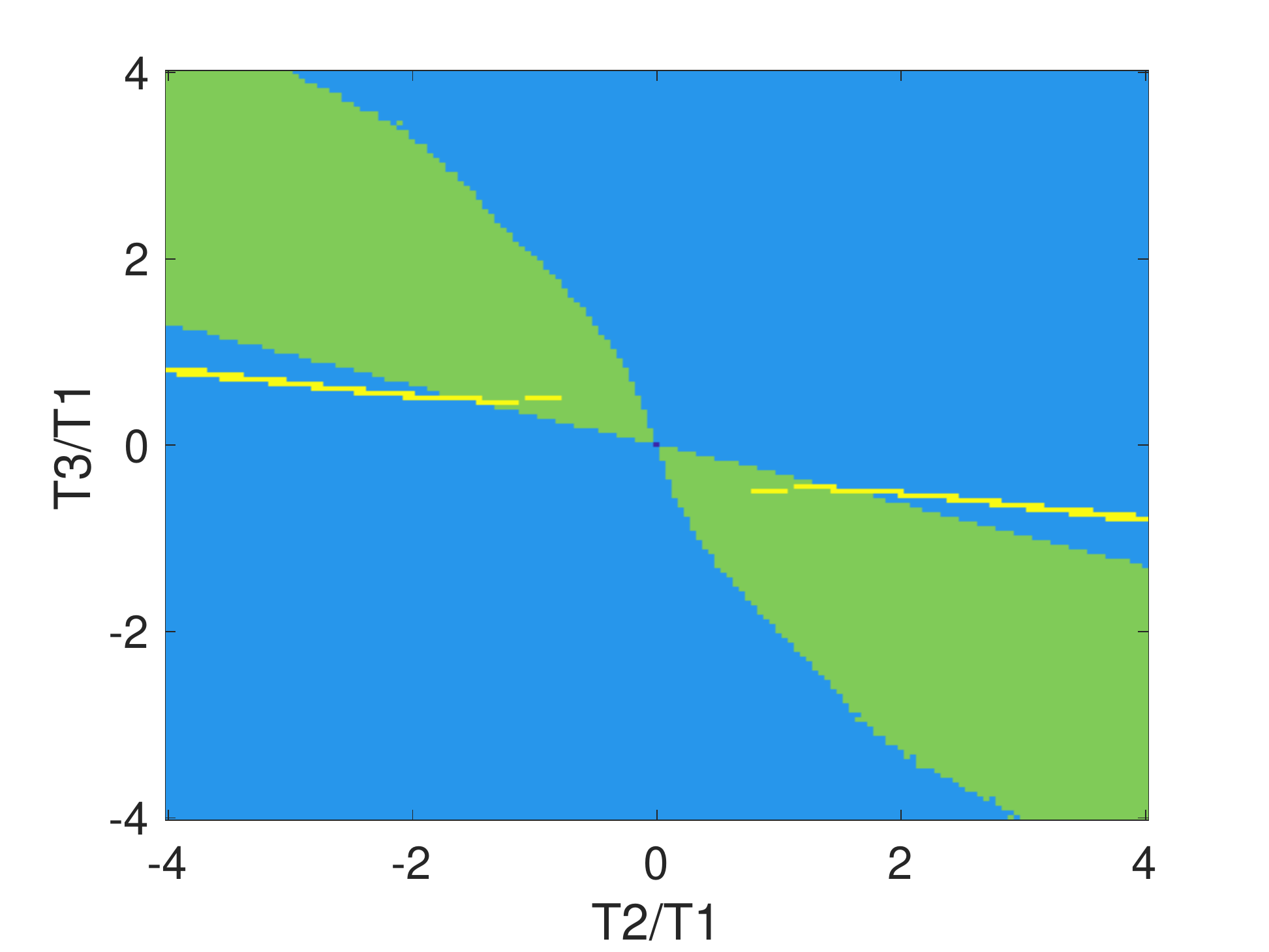} 
\par\end{centering}
\caption{\label{fig:sliver} (color online) s-wave (blue) vs d-wave (green)
phase diagram of the mean field fermionic dimer model in presence
of antiferromagnetic interaction $J/T_{1}=50$, where we have chosen
$T_{1}>0$. %For $T_{2}=T_{3}=0$ the two instabilities have the same
%free energy (dark blue). 
Highlighted in yellow is the locus of the points in the $T_{2}/T_{1},\,T_{3}/T_{1}$
plane where the van Hove singularity occurs for $p\in(0.18,0.22)$.
Notice the small but finite overlap with the d-wave region, near $T_{2}/T_{1}\simeq\mp0.8$
and $T_{3}/T_{1}\simeq\pm0.5$.}
\end{figure}

We find that the model is able to satisfy the conditions (i) and (ii)
in a small sliver in the $T_{2}/T_{1},\,T_{3}/T_{1}$ plane (see Fig.~\ref{fig:sliver}),
only if we choose the sign of $T_{1}$ to be positive. 
In order to assess whether any portion of the identified sliver is
consistent with the model exhibiting d-wave superconductivity, we
consider the effect of the antiferromagnetic interaction $J$ of the
$t-J$ model from which the FB model descends, and follow the procedure
in Ref.~\onlinecite{Goldstein2017} to compare s-wave vs d-wave free
energies. The choice of value of the ratio $J/T_{1}$ of interaction
strength $J$ to the scale $T_{1}$ is non trivial, since in our phenomenological
approach we do not determine $T_{1}$ from first principles. (The
value of $T_{1}$ can be much smaller than the value of $t$ because
of the suppression coming from the condensation of the bosonic dimers.)
However, we find as it is reasonable to expect that the phase boundaries
between \textit{s}-wave and \textit{d}-wave as a function of system
parameters become independent of $J$ when $J/T_{1}\gg1$, and the
latter in general favors d-wave superconductivity. For this reason
we opted to work in the large $J/T_{1}$ limit and thus obtain an
upper bound to the portion of parameter space where (i), (ii), and
(iii) are satisfied. This is illustrated in Fig.~\ref{fig:sliver}
by the overlap between the sliver and the d-wave portion of the phase
diagram (shown for $J=50$ in units of $T_{1}$). What we find is
a narrow but non-vanishing region in parameter space, located around
$T_{2}/T_{1}=\mp0.8$ and $T_{3}/T_{1}=\pm0.5$. The dispersion of
the system at these points is shown in the top panel in Fig.~\ref{fig-vanHove}.

When a van Hove singularity occurs in a 2D fermionic systems, the
effective mass of the quasiparticle excitations diverges logarithmically
as 
\begin{equation}
m^{*}\sim\,a\;\log\frac{\Lambda}{|\varepsilon|}\,,\label{eq:Mass_divergence}
\end{equation}
where $\varepsilon$ is the energy from the van Hove singularity,
$\Lambda$ is the bandwidth, and $a$ has dimensions of mass. The
inverse mass as a function of $p$ is also shown for $T_{2}/T_{1}=-0.8$
and $T_{3}/T_{1}=0.5$ in Fig.~\ref{fig-data}. Near the van Hove
singularity, it is possible to obtain a perturbative analytical expression
that relates the inverse mass to the doping, 
\begin{equation}
\vert p-p_{c}\vert=b\left(\frac{m^{*}}{a}+1\right)\,e^{-m^{*}/a}\,,\label{eq:Mass_divergence pert}
\end{equation}
where $a\simeq0.25$ and $b\simeq0.38$ are found most conveniently
by fitting to the numerical data (thin line near $p_{c}$ in Fig.~\ref{fig-data}).

We further checked that d-wave is the leading superconducting instability
for these values of $T_{2}/T_{1},\,T_{3}/T_{1}$, for a range of values
of $J/T_{1}$ (see Appendix \ref{subsec:Superconducting-gap-and}).
We find that the superconducting gap $\Delta$ scales as $\Delta\sim0.02J$.
In Fig.~\ref{fig-data} we show the value of the superconducting
gap for $J=1.0$ in units of $T_{1}$. This choice takes into account
that the ratio $J/t\sim0.2-0.4$, and that $T_{1}/t$ is similarly
suppressed with respect to $t$.

As discussed above, and illustrated in Fig.~\ref{fig-vanHove}, the
van Hove singularity considered here separates a region with four
identical Fermi pockets of size $p/8$ from a region with two (much
larger) Fermi surfaces, with one surface encasing the other. We therefore
expect two distinct features as the system crosses the singularity:
(a) a discontinuous jump in the period of oscillations; and (b) the
appearance of two (much smaller) distinct periods for $p>p_{c}$.
However, it may well happen that the experimental validity of the
FB dimer model does not extend to the overdoped regime and breaks
down at $p_{c}$. Further work beyond the scope of the present paper
is needed to ascertain this possibility and investigate alternative
scenarios as $p$ is tuned across the singularity.

\section{\label{sec:Conclusions}Conclusions}

In this work we studied quantum oscillations in a FB dimer model for
high temperature superconductors, within a mean field approximation.
We argued that our system satisfies Onsager quantization and the Lifshitz-Kosevich
formula. We studied the effective mass for quantum oscillations and
found that it diverges at a van Hove singularity, where four Fermi
pockets merge pairwise at a critical doping at four different points
in the Brillouin zone. The location of the singularity depends on
the effective fermionic dimer hopping parameters, and we narrowed
down the range of such parameters for the model to contain pockets
in the underdoped regime, display d-wave superconductivity, and have
the singularity near optimal doping $p\sim0.2$.

We find that we can match rather well the experimental quantum oscillation
behavior in the cuprates. This is remarkable given the simplicity
of the effective dimer model. It is furthermore enticing that the
agreement occurs for a relatively narrow range in parameter space;
our results bring us closer to proposing a comparison of the behavior
of the mean field dimer model with experiments that will critically
ascertain its limits of validity.

One of the predictions we make is that across the van Hove singularity
the quantum oscillation frequency jumps discontinuously to much larger
values and two periods appear. The current state-of-the-art high-field
capability does not allow one to study quantum oscillations near optimum
doping in the cuprates to verify this prediction. However, it may
well be within range of near future improvements in the experimental
technique.

The enhanced density of states at the van Hove singularity consistently
coincides with a maximum in the superconducting gap at the value of
doping corresponding to that where the Fermi surface topology changes.
This result supports a theoretical proposal of a quantum critical
point near optimum doping associated with a van Hove singularity where
four Fermi pockets merge inside the Brillouin zone (not at its boundary).
Figure~\ref{fig-data} highlights our proposed scenario. We expect
this finding to have observable consequences in the quantum critical
region, for instance on the temperature dependence of the resistivity.
Our results thus give a concrete motivation to study quantum criticality
at a van Hove singularity.

\textit{Acknowledgements.} We are very grateful to Nigel Cooper for
many useful discussions that helped us shape this project and understand
the nature of quantum oscillations in the fermion boson dimer model.
This work was supported, in part, by the Engineering and Physical
Sciences Research Council (EPSRC) Grant No. EP/M007065/1 (C.Ca. and
G.G.), and by DOE Grant No. DE-FG02- 06ER46316 (C.Ch.). Statement
of compliance with the EPSRC policy framework on research data: this
publication reports theoretical work that does not require supporting
research data.

\appendix
%dummy comment inserted by tex2lyx to ensure that this paragraph is not empty

\section{\label{subsec:Superconducting-gap-and}Superconducting gap and inverse
mass}

\begin{figure}[H]
\begin{centering}
\includegraphics[width=0.9\columnwidth]{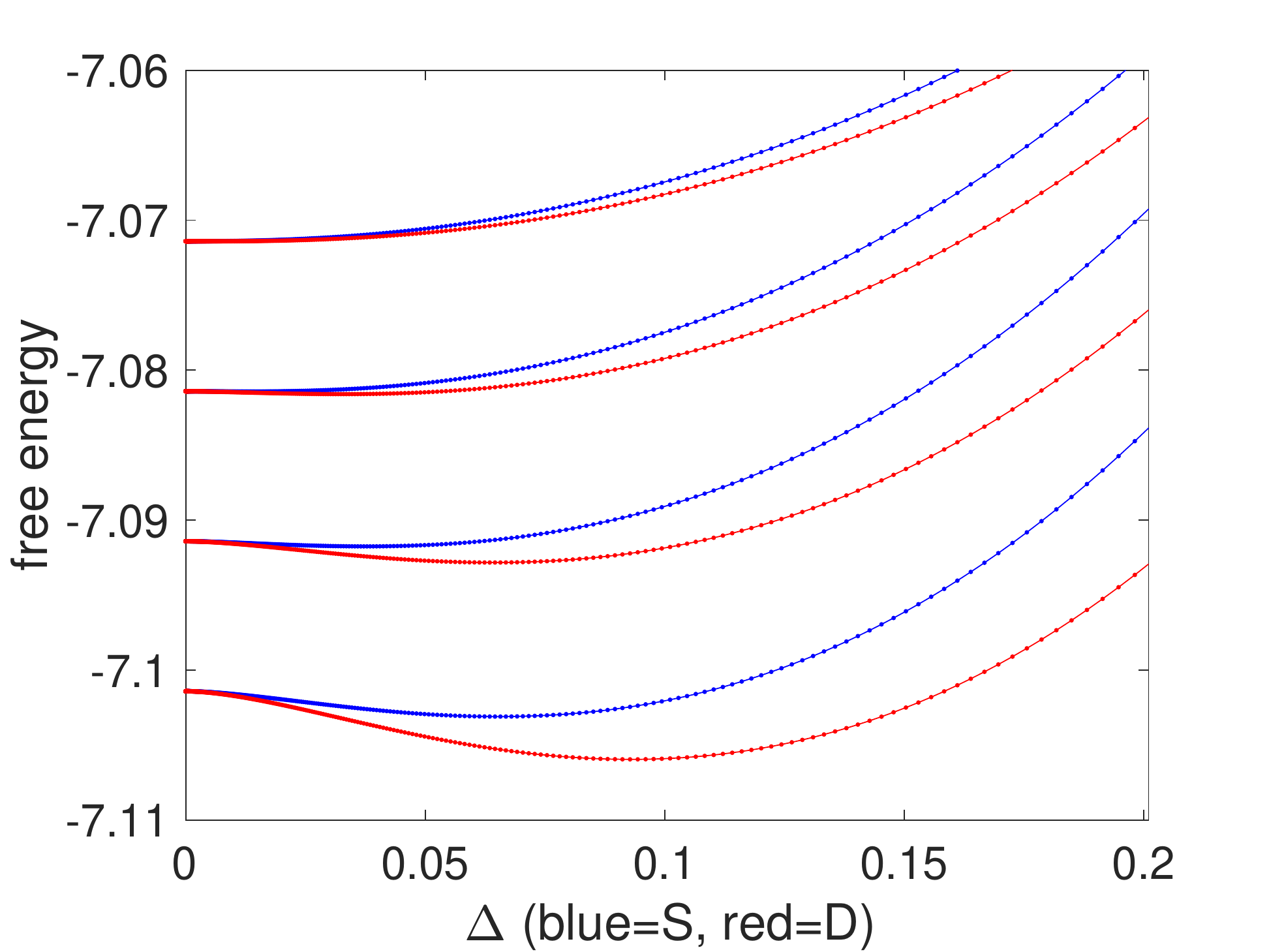} 
\par\end{centering}
\caption{\label{fig-Jcheck} (color online) Comparison of s-wave (blue) vs
d-wave (red) free energies as a function of the gap $\Delta$ for
$T_{2}/T_{1}=-0.8$ and $T_{3}/T_{1}=0.5$, for $J/T_{1}=1.5,2.5,3.5,4.5$
(top to bottom pairs of curves). The value of the chemical potential
was chosen near the van Hove saddle point. An artificial offset has
been introduced (with respect to the bottom pair of curves) for visualization
purposes. Without offset, all the curves coincide at $\Delta=0$.
In all cases (although it is difficult to see for small values of
$J$ in the figure), the minimum of the \textit{d}-wave free energy
occurs at a finite value of $\Delta$ and is lower than the minimum
of the \textit{s}-wave free energy.}
\end{figure}

\begin{figure}[H]
\begin{centering}
\includegraphics[width=0.9\columnwidth]{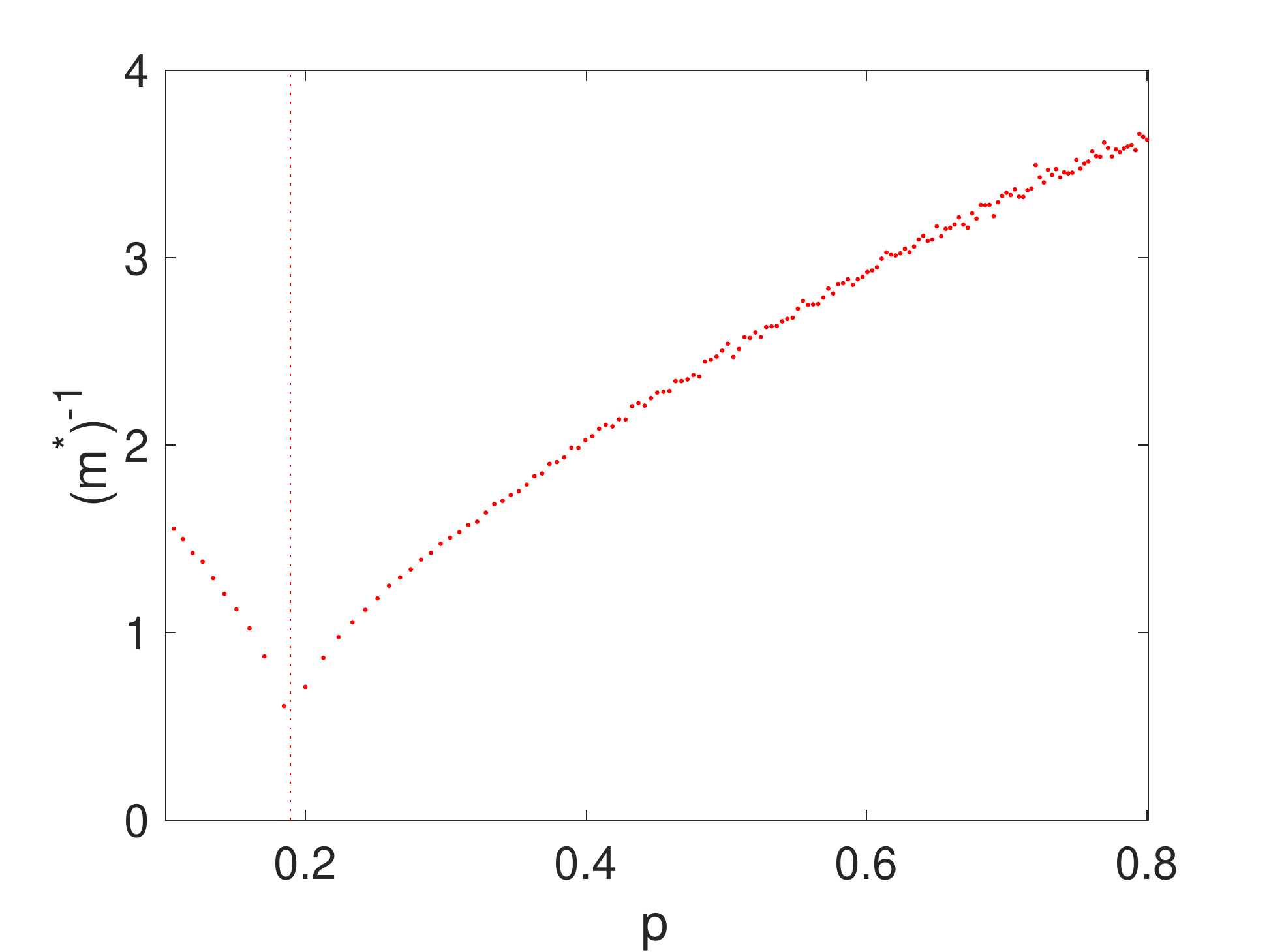} 
\par\end{centering}
\caption{\label{fig-invmassVSp} Inverse quasiparticle mass for $p\in(0,0.8)$
at $T_{2}/T_{1}=-0.8$ and $T_{3}/T_{1}=0.5$. The vertical dotted
line indicates the position of the van Hove singularity considered
in the main text. (These are the same data shown in Fig.~1 of the
main text for a narrower range of doping.)}
\end{figure}

\begin{figure}[H]
\begin{centering}
\includegraphics[width=0.9\columnwidth]{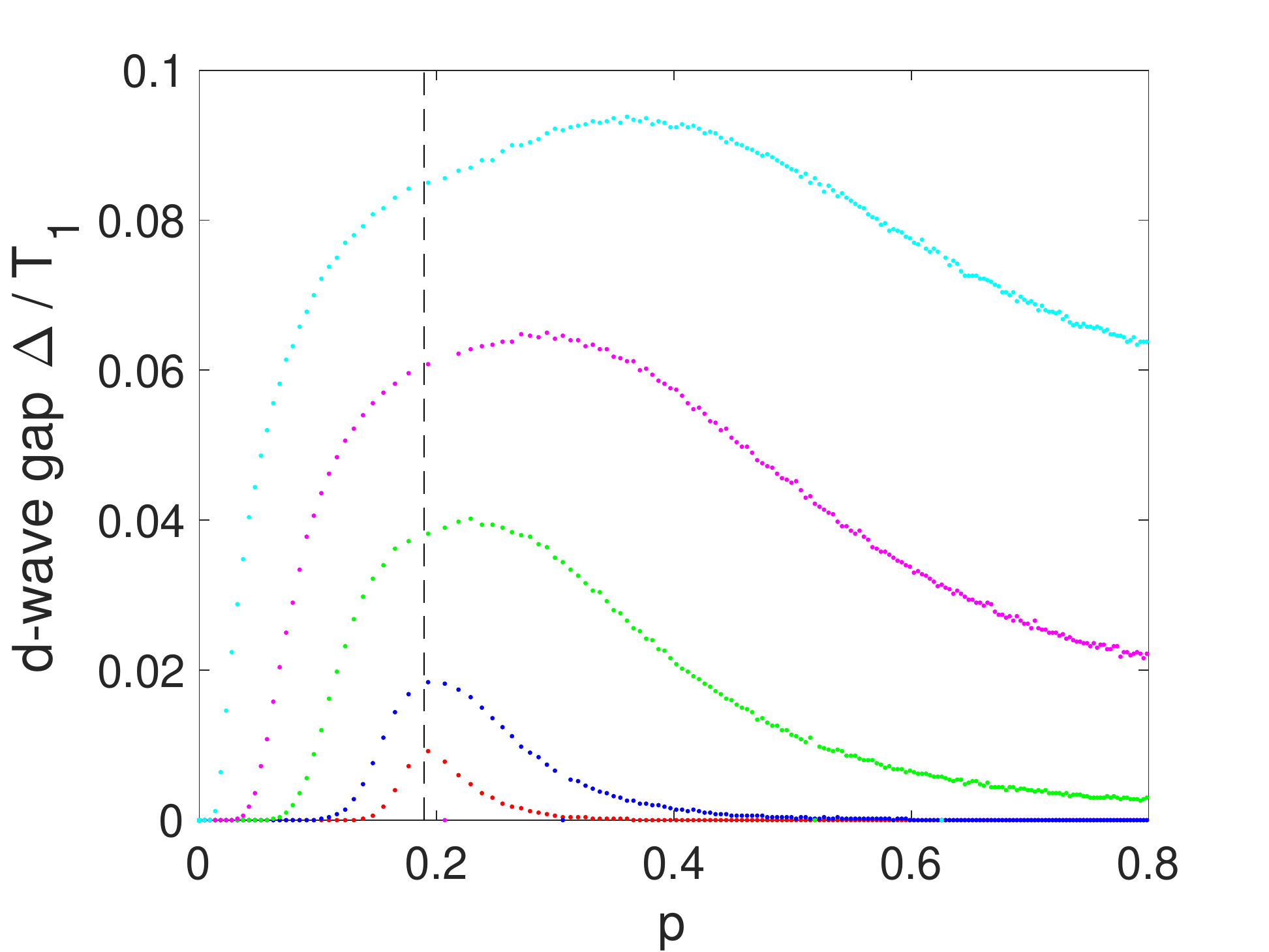} 
\par\end{centering}
\caption{\label{fig-gapVSp} (color online) d-wave gap $\Delta/T_{1}$ for
$T_{2}/T_{1}=-0.8$, $T_{3}/T_{1}=0.5$ and $J/T_{1}=1.0,1.5,2.5,3.5,4.5$
(red, blue, green, magenta, cyan). The vertical dashed line indicates
the location of the van Hove singularity.}
\end{figure}

In Fig.~\ref{fig-Jcheck} we verify that the d-wave superconducting
instability is the leading instability for $T_{2}/T_{1}=-0.8$ and
$T_{3}/T_{1}=0.5$, for a range of values of the interaction $J$.
The mean field dimer model ceases to be a good representation of the
original FB dimer model, and even more so of the underlying electronic
system, when the density of fermionic dimers $p$ increases. With
this caveat in mind, we show for completeness in Fig.~\ref{fig-invmassVSp}
the behavior of the inverse mass of the mean field dimer model over
a larger interval in $p$. Further van Hove singularities occur for
$p>0.8$ (not shown). In Fig.~\ref{fig-gapVSp} we then show the
behavior of the \textit{d}-wave gap on the broader range of $p$,
for different values of $J$. We note that larger values of $J$ tend
to mix the small $p$ behavior with the large $p$ behavior of the
model (namely, other van Hove singularities for $p>0.8$) and therefore
they ought to be considered with care.

From Fig.~\ref{fig-gapVSp} we observe that the dependence of the
maximum value of $\Delta(p)/T_{1}$ on the coupling strength $J/T_{1}$
is approximately linear, following the relation $\Delta\sim0.02J-0.01$
reported in the main text.

\section{\label{sec:Frequency-of-the}Frequency of the oscillations}

\begin{figure}
\begin{centering}
\includegraphics[width=1\columnwidth]{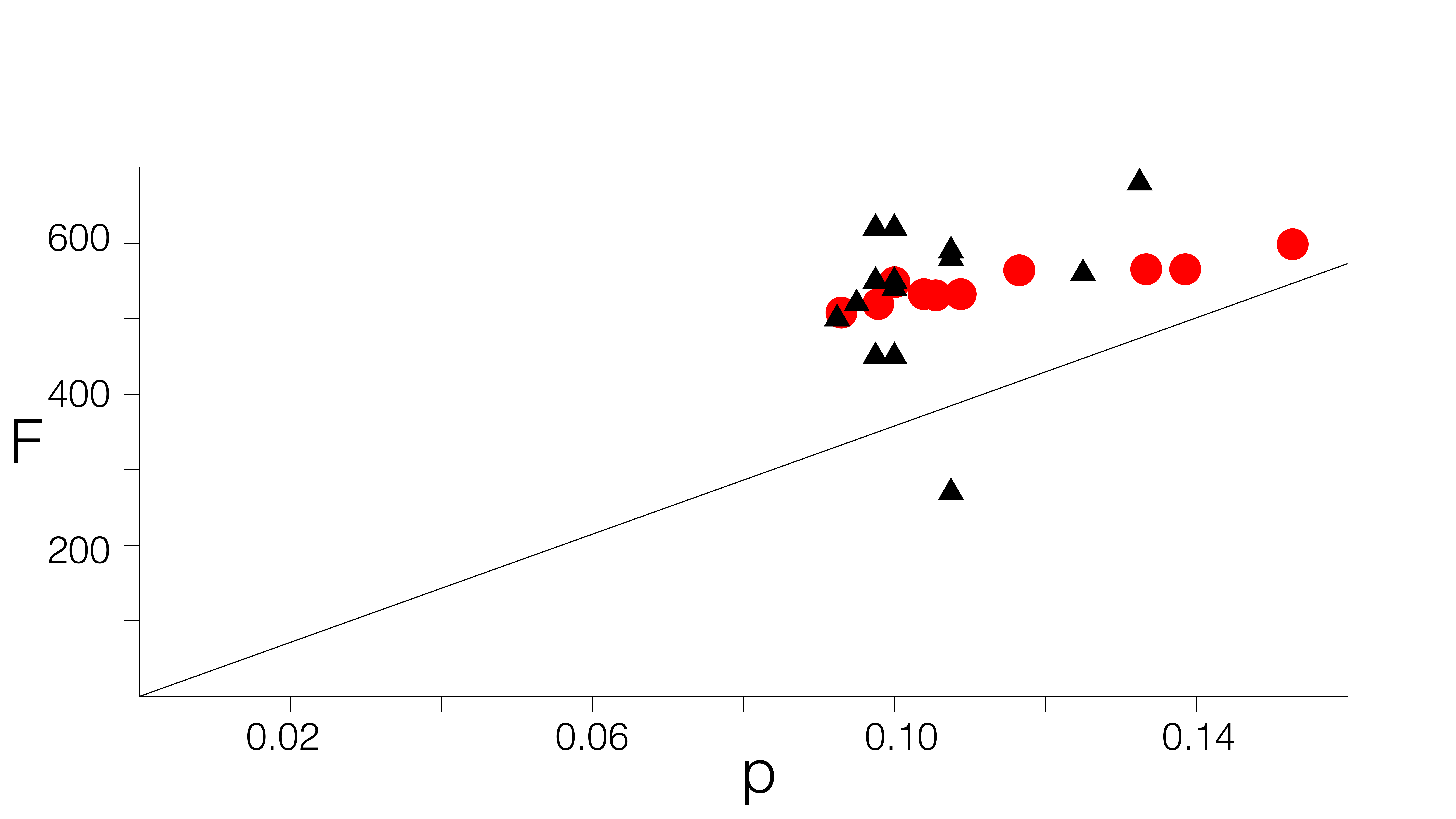} 
\par\end{centering}
\caption{\label{fig:Fit} Frequency $F$ of the quantum oscillations (in Tesla)
as a function of doping $p$ from two independent experimental groups
(Ref.~\protect\protect\onlinecite{Singleton2010} in red and Ref.~\protect\protect\onlinecite{Ramshaw2015}
in black). The black solid line is Eq.~\eqref{eq:Oscillations}.}
\end{figure}

If we assume a Fermi surface of area $p/8$ and a lattice constant
of $3.8$~Å, we find that the frequency of oscillations $F$ (measured
in Tesla) is related to the doping $p$ in the cuprates as: 
\begin{equation}
F=3.58\cdot10^{3}\cdot p\,.\label{eq:Oscillations}
\end{equation}
According to this relation, a typical frequency of $530$~T corresponds
to $p=0.148$. Equation~\eqref{eq:Oscillations} is compared to experimental
quantum oscillations data from Ramshaw \textit{et al.}~\cite{Ramshaw2015}
(red dots) and from Singleton \textit{et al.}~\cite{Singleton2010}
(black triangles) in Fig.~\ref{fig:Fit}. 
%\begin{thebibliography}{10}

%\end{thebibliography}

%%%%%%%%%%%%%%%%%%%%%%%%%%%%%%%%%%%%%%%%%%%%%%%%%%%%%%%%%%%%%%%%%%%%%%%
%%%%%%%%%%%%%%%%%%%%%%%%%%%%%%%%%%%%%%%%%%%%%%%%%%%%%%%%%%%%%%%%%%%%%%%

%%%%%%%%%%%%%%%%%%%%%%%%%%%%%%%%%%%%%%%%%%%%%%%%%%%%%%%%%%%%%%%%%%%%%%%

\end{document}